\documentclass[journal]{IEEEtran}
\IEEEoverridecommandlockouts
\usepackage{flushend}
\usepackage{amsfonts}
\usepackage{textcomp}
\usepackage{xcolor}
\usepackage{fancyhdr}
\usepackage{cite}
\usepackage{graphicx}
\usepackage{psfrag}
\usepackage{subfigure}
\usepackage{url}
\usepackage{stfloats}
\usepackage{amsmath}
\usepackage{array}
\usepackage{fancyhdr}
\usepackage{epsfig}
\usepackage{amssymb}
\usepackage{color}
\usepackage{float}
\usepackage{algorithm}
\usepackage{algpseudocode}
\usepackage{balance}
\usepackage{soul}
\usepackage{caption}
\usepackage{amsthm}
\usepackage{balance}
\newtheorem{definition}{Definition}

\title{Cognitive Semantic Augmentation LEO Satellite Networks for Earth Observation}
\author{\IEEEauthorblockN{Hong-fu Chou, Vu Nguyen Ha,  Prabhu Thiruvasagam, Thanh-Dung Le, Geoffrey Eappen, Ti Ti Nguyen, Duc Dung Tran, Luis M. Garces-Socarras, Juan Carlos Merlano-Duncan, Symeon Chatzinotas\\}
\IEEEauthorblockA{\textit{Interdisciplinary Centre for Security, Reliability and Trust (SnT), University of Luxembourg, Luxembourg}}}

\begin{document}
\maketitle

\begin{abstract}
Earth observation (EO) systems are essential for mapping, catastrophe monitoring, and resource management, but they have trouble processing and sending large amounts of EO data efficiently, especially for specialized applications like agriculture and real-time disaster response.  This paper presents a novel framework for semantic communication in EO satellite networks, aimed at enhancing data transmission efficiency and system performance through cognitive processing techniques. The proposed system leverages 
Discrete Task-Oriented Joint Source-Channel Coding (DT-JSCC) and Semantic Data Augmentation (SA) integrate cognitive semantic processing with inter-satellite links, enabling efficient analysis and transmission of multispectral imagery for improved object detection, pattern recognition, and real-time decision-making. Cognitive Semantic Augmentation (CSA) is introduced to enhance a system's capability to process and transmit semantic information, improving feature prioritization, consistency, and adaptation to changing communication and application needs. The end-to-end architecture is designed for next-generation satellite networks, such as those supporting 6G, demonstrating significant improvements in fewer communication rounds and better accuracy over federated learning. 
\end{abstract}
\section{Introduction}
The United Nations has acknowledged Earth observation (EO) as a crucial instrument for accomplishing many Sustainable Development Goals. EO systems acquire and analyze data about the globe for numerous civil applications and critical tasks that significantly impact human existence through the use of satellite networks. These tasks include mapping, weather forecasting, risk management, disaster monitoring, natural resource management, and emergency response. Semantic image processing, a key task in computer vision, involves assigning distinct labels to each pixel in an image. This pixel-level categorization is crucial for numerous Earth Observation (EO) applications, such as land cover classification, disaster response, and environmental monitoring. In these scenarios, satellite imagery must be interpreted with precision. However, challenges like occluded objects, complex environments, and cluttered backgrounds can complicate the process. Fortunately, recent advances in deep learning have greatly enhanced the performance of semantic segmentation, making it an increasingly effective tool for analyzing EO data.

The rapid growth in data generated by satellites and other remote sensing technologies presents new challenges. While this data holds enormous potential for tasks such as environmental monitoring and disaster relief, its sheer volume makes processing and interpretation difficult. In this context, \cite{li2021image} explores modern methods for analyzing and managing large-scale EO datasets. Leveraging advancements in computer vision, machine learning, and interdisciplinary approaches, the study underscores the importance of semantic knowledge in improving the efficiency and accuracy of EO data analysis.

An innovative approach to addressing these challenges is the CORSA system \cite{ivashkovych2022corsa}, which introduces an AI-driven compression technique that preserves the integrity of data while producing efficient representations. CORSA demonstrates its ability to compress EO data, such as water detection networks, without sacrificing accuracy, highlighting its adaptability for on-board satellite processing. This method provides a powerful solution to the growing demands of managing and analyzing vast amounts of EO data. Another significant development in EO data processing is the SAMRS framework \cite{wang2024samrs}, which combines object detection techniques with the Segment Anything Model (SAM) to perform effective semantic labeling and segmentation. SAMRS utilizes a vast dataset of over 100,000 images with precise EO data, making it a versatile tool for various semantic segmentation tasks across diverse urban landscapes. Incorporating semantic knowledge to simplify trajectory data represents a further breakthrough in EO processing. The work in \cite{liu2021semantics} presents an innovative method that enhances trajectory simplification by distinguishing periods of movement from stationary phases within trajectories. This refined segmentation results in more meaningful representations of movement patterns, improving the analysis and interpretation of trajectory data in EO contexts.
\begin{table*}[h]
\centering
\caption{Summary of Methodologies and Applications in Semantic Image Processing and Compression}
\scriptsize
\begin{tabular}{|c|c|c|c|c|c|c|c|}
\hline
\textbf{Feature} & \textbf{\cite{ivashkovych2022corsa}} & \textbf{\cite{wang2024samrs}} & \textbf{\cite{liu2021semantics}} & \textbf{\cite{hong2023cross}} & \textbf{\cite{weixiao2023pssnet}} & \textbf{Proposed} \\ \hline
EO data processing & \checkmark & \checkmark & \checkmark & & \checkmark& \checkmark \\ \hline
AI-based compression & \checkmark & & \checkmark & \checkmark & & \checkmark \\ \hline
Semantic labeling & & \checkmark & & & \checkmark&  \\ \hline
Remote sensing & & \checkmark & & \checkmark & \checkmark & \checkmark \\ \hline
On-board satellite processing & \checkmark & & \checkmark & \checkmark & & \checkmark \\ \hline
Multi-spectral Analysis & \checkmark & \checkmark & & & & \\ \hline
\end{tabular}
\label{table1}
\end{table*}

The HighDAN AI network, detailed in \cite{hong2023cross}, extends these capabilities to urban environments by applying deep learning for semantic compression. HighDAN excels at interpreting complex EO imagery, such as satellite images, across different urban settings. Its ability to generalize effectively across diverse geographical contexts makes it an invaluable tool for urban geospatial analysis. The exploration of deep learning for semantic segmentation in three-dimensional meshes, as presented in \cite{weixiao2023pssnet}, marks another important frontier in EO data analysis. This study applies deep learning techniques to 3D models, focusing on the classification of structures like urban buildings and roadways. By advancing the segmentation of 3D meshes, this research expands the applicability of semantic segmentation beyond traditional 2D satellite imagery, offering new insights for complex EO datasets.

Table \ref{table1} highlights the significant advancements in semantic image processing, computational semantic compression, and the fusion of semantic trajectories within the context of EO and related fields. From streamlining data searches and enhancing data integration to innovative compression methods like CORSA, these contributions emphasize the growing importance of semantic technologies in handling large-scale EO datasets. The development of semantic EO data cubes, real-time semantic trajectory processing systems, and advanced AI models such as HighDAN and SAMRS demonstrates how semantic understanding is transforming data analysis, enabling more efficient and accurate results. These innovations pave the way for improved applications across various domains, from autonomous vehicles and urban planning to medical image analysis and video surveillance. Nonetheless, challenges remain, particularly with respect to non-standard data types and limited datasets, requiring further research to optimize deep learning models and enhance generalizability across diverse environments.

We summarize our contributions as follows:
\begin{itemize}
    \item The cognitive semantic EO system model enhances satellite networks by integrating DT-JSCC \cite{xie2023robust} and Semantic Data Augmentation (SA) \cite{pu2024fine} techniques. These techniques enable the transmission of task-relevant semantic features, rather than raw data, and leverage the semantic structure of EO data to improve the inference accuracy at the receiver side.
    \item The framework integrates inter-satellite communication and deep learning models for on-board processing, enabling an end-to-end semantic-augmented communication process. This optimizes the transmission, analysis, and decision-making of multispectral EO data, resulting in improved classification accuracy and overall system efficiency.
    \item Through the use of semantic communication, this approach markedly has the potential to reduce communication overhead and energy consumption by using a lower complexity DTJSCC framework, meeting the efficiency demands of 6G satellite communication.    
\end{itemize}

\section{Formation of semantic communication networks}
The utilization of semantic communication (SemCom), which specifically prioritizes the transmission of the fundamental meaning (semantics) of information, has been increasingly popular because of its ability to create more efficient communication systems. Notwithstanding this commitment, there remains a discrepancy between theoretical principles and their practical execution. 
\subsubsection{Orientation and attributes of semantic communication}
As Shannon's Law nears its ultimate limits, researchers are exploring the theoretical constraints of compressing substantial information (semantic information). Through the use of this semantic source, the authors have derived the theoretical limits for both lossless and lossy compression, along with the minimum and maximum constraints on the rate-distortion function.  The primary factor for this is its ability to collect and compress the most pertinent and concise information, and then summarize it effectively.  
\begin{itemize}
    \item Evaluation of the level of compression achieved in semantic source coding with respect to the original information is crucial. In order to determine the theoretical boundaries for lossless and lossy compression. Moreover, they established the lower and upper limits for the rate-distortion function.
    \item The aim of the task-oriented component is to effectively accomplish tasks while taking into account the possible influence on the value of the information delivered \cite{mostaani2022task}. Leveraging easily accessible resources such as communication bandwidth, processing expenses, and energy consumption enhances the feasibility of attaining a shared goal within the constraints and specifications of a particular project. 
    \item An assessment of the effectiveness of a task or performance indicator relies on the extent to which the obtained information contributes to its achievement. System performance evaluation may be quantified by measuring the extent to which a certain objective is achieved, taking into account the allocated resources.
\end{itemize}
These previously described perspectives differ from the approach grounded on the information theory paradigm, which considers all potential transmission sequences.

\subsubsection{Semantic joint source-channel coding}

The partnership between JSCC and deep learning showcases that the deep learning encoder and decoder outperform the traditional approach in terms of word error rate, particularly when there are limitations on the computational resources designated for syllable encoding. A fundamental limitation of this approach is the use of a predetermined bit size to encode phrases with varying lengths. The primary goal of deep learning-based Joint Source-Channel Coding (DJSCC) is to reliably attain high levels of performance, even in scenarios characterized by resource constraints and low Signal-to-Noise Ratios (SNRs).

Two innovative approaches are proposed in the research of Digital Task-oriented Joint Source-Channel Coding (DT-JSCC) \cite{xie2023robust} to facilitate the seamless integration of digital communication and DJSCC. The first modulation, uniform modulation, functions as a flexible spring by modifying the spacing between data points (constellations) during transmission to align more effectively with the unique properties of the picture data being sent.  The potential advantages outlined in \cite{Zhong2024} of incorporating Unequal Error Protection (UEP) into a semantic encoder/decoder, using the current JSCC system, are worth investigating. The objective of this approach is to preserve the relative importance of data focused on semantic tasks. Furthermore, the implementation of a JSCC system may be simplified by using Quasi-cyclic Low-Density Parity-Check (QC-LDPC) codes on an adaptable device. A prototype for semantic communications is developed by constructing a fixed-point system that is then used to transmit and receive semantic information across a channel that contains residual noise. 

\section{On-board semantic feature processing and deep learning for Resource-efficient EO systems}
This section introduces a cognitive semantic EO system model designed to enhance satellite communication networks by integrating advanced semantic processing techniques, such as DT-JSCC  and SA techniques. The model leverages inter-satellite links and cognitive semantic augmentation to optimize the transmission and analysis of multispectral EO data. By focusing on transmitting essential, task-relevant semantic features rather than raw data, the system reduces communication overhead and energy consumption, while improving decision-making and performance in 6G satellite communication networks. The proposed framework showcases an end-to-end semantic-augmented communication process, ensuring efficient data handling and enhanced user experience. Furthermore, deep learning models for on-board processing are revisited based on complexity and efficiency, focusing on metrics like model size, parameters, and training/inference time. 
\subsection{Cognitive Semantic EO system model}
\begin{figure*}[htbp]
\centerline{\includegraphics[width=\textwidth]{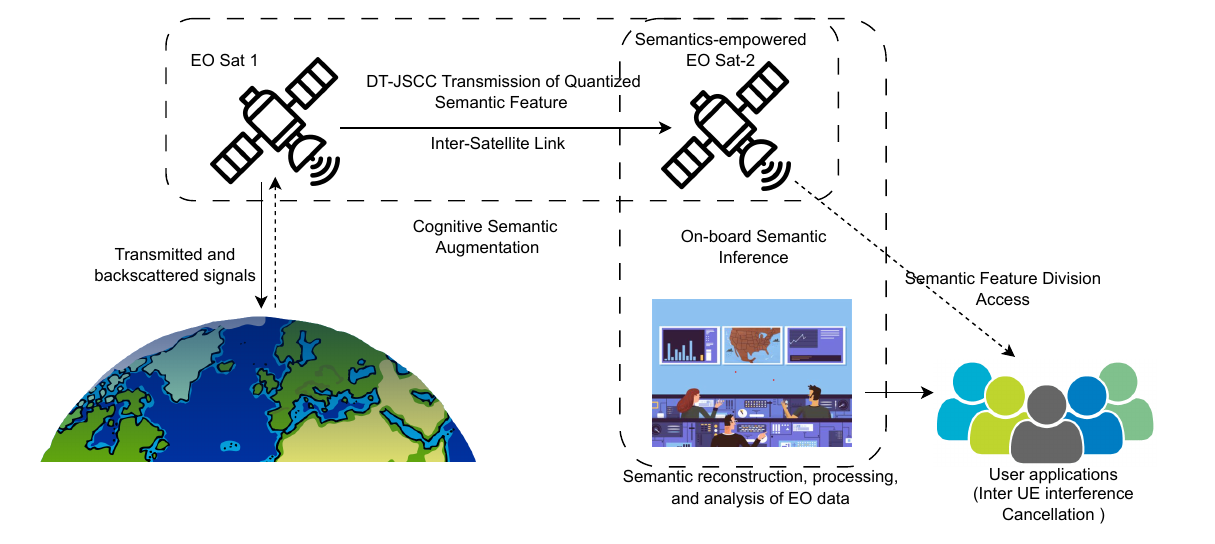}}
\caption{The proposed cognitive semantic augmentation LEO satellites for enhanced EO systems.}
\label{2SatSEM}
\end{figure*}
In this satellite system illustrated in Fig.\ref{2SatSEM}, a low Earth orbit (LEO) satellite establishes communication with a gateway located at a ground terminal directly connected to a server. Moreover, the satellite is smoothly integrated into an intricate network of inter-satellite links, allowing it to exchange real-time data with other spacecraft in orbit. Implementing an inter-satellite connection improves data transmission reliability and extends the communication period before the satellite moves outside the ground terminal's coverage area.

The system processes multispectral satellite images captured at different time points, utilizing two satellites. The Sat 1, positioned near the ground terminal, captures the reference image at time $t_0$, denoted as $U(t_0)$. This image is represented as $
U(t_0) = \{u^{t_0}(i, j, k) \mid 1 \leq i \leq H, 1 \leq j \leq W, 1 \leq k \leq D\}$, where $H$ is the height, $W$ is the width, and $D$ is the number of spectral bands.

\subsubsection{Semantic data transmission} Upon cognitive processing of the reference image $U(t_0)$, semantic information $U_{sem}(t_0)$ is generated by the
classification network $\textit{f}_{s1}$ of on-board processor on Sat 1. This semantic information can be mathematically represented as $U_{sem}(t_0)=\textit{f}_{s1}(U(t_0))$, where $U_{sem}(t_0)=\{u^{t_0}(i, j, k) \mid 1 \leq i \leq B, 1 \leq j \leq C, 1 \leq k \leq A\}$. We define $B$ as the batch size, $C$ as the number of classes, and $A$ as the feature dimension. This information is then sent across an inter-satellite communication connection to the second satellite, which is located at a further distance and serves as a relay. The transmission and reception process can be executed using the DT-JSCC approach \cite{xie2023robust}, which utilizes a defined quantization level to incorporate error correction for channel coding. A parallel can be drawn to the collaborative efforts of the European Space Agency's Sentinel satellites and NASA's Landsat program in observing the Earth. While Sentinel provides distinct spectral capabilities and more frequent revisit cycles, Landsat contributes long-term, high-resolution imagery. Together, these satellites monitor changes in agricultural practices, urbanization, deforestation, and land usage.

The integration of this shared data enables researchers and policymakers to more effectively manage natural resources, track environmental changes, and formulate strategies to combat climate change and protect biodiversity. As illustrated in Fig.\ref{SEMtxrx}, at time $t_1$, Sat 2 leverages the semantic information $U_{sem}(t_0)$ transmitted by Sat 1, encoded as $\textit{f}_{s2}$, to establish a correlation with the current image $U(t_1)$ that Sat 2 has obtained. This process can be refined through on-board training to enhance the extraction of semantic features. This ensures that Sat 2 can analyze the new image effectively by referencing the information previously received from Sat 1. The cognitive semantic channel indicates the
learning parameter modified by the covariance prediction net-
work during the execution of end-to-end semantic-augmented
communication discussed in the subsequent sections.

\subsubsection{Cognitive semantic augmentation}
\begin{figure*}[htbp]
\centerline{\includegraphics[width=\textwidth]{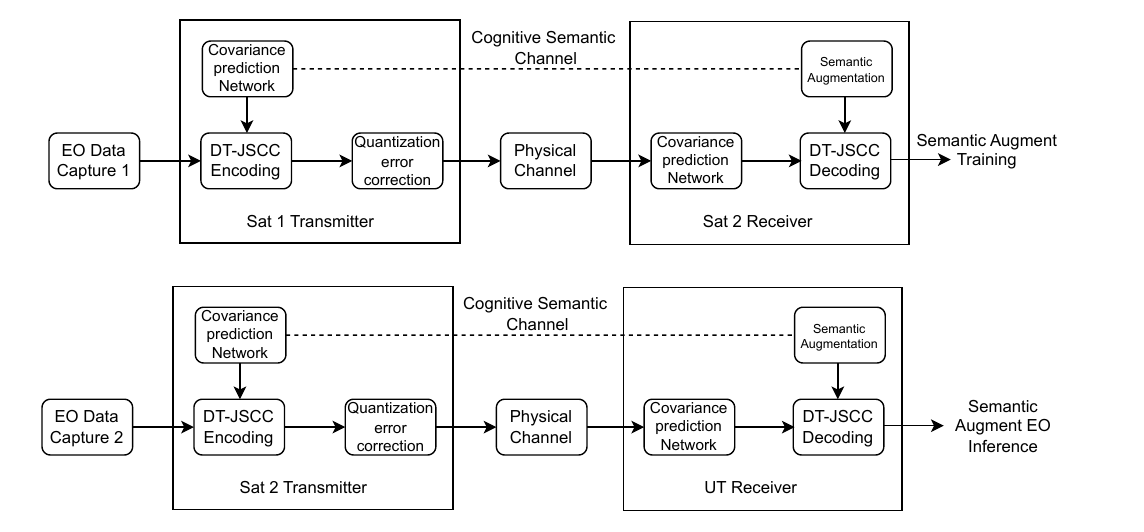}}
\caption{The proposed transmitter and receiver for cognitive semantic satellite networks.}
\label{SEMtxrx}
\end{figure*}

Sat 2 utilizes semantic information $U_{sem}(t_0)$ to extract and acquire pertinent features from the $U(t_1)$ image. Semantic cognitive augmentation intends to prioritize both the pertinent data for the task and the semantic feature invariant, facilitating the effective detection of important changes and correlations between the reference and new images through the SA methodology \cite{pu2024fine}.  As illustrated in Fig.\ref{SEMtxrx}, the covariance matrix prediction network $\textit{g}_{s2}$ learns from $U_{sem}(t_0)$ to minimize the SA training loss $\textit{L}_{SA}(\textit{f}_{s2};\textit{g}_{s2}|U_{sem}(t_0))$. This can be done by setting up an SA task and optimizing $\textit{g}_{s2}$ by backpropagating the SA-gradient $\theta_{\textit{g}_{s2}}$ and minimizing the cross-entropy loss of the output of DT-JSCC decoder $\textit{l}_{s2}$. Processing data on the semantic feature domain $U_{sem}(t)$, as opposed to the multi-spectral pixels domain $MP(t)$, shows potential for transmitting cognitive information effectively to satellites in close proximity. We provide the semantic cognitive covariance matrix $\upsilon_{t_1}^{\textit{g}_{s2}} = \textit{g}_{s2}(U_{sem}(t_0),\theta_{\textit{g}_{s2}})$, which enhances the generalization of image change in EO data and facilitates a smooth integration into the optimization objective of the classification network. The scope of this extends beyond particular tasks linked to the development of certain neural networks. Therefore, the cognitive semantic channel can be facilitated by the above methodology of integrating DT-JSCC and semantic augmentation technique, which expands the applications of classification tasks for 6G communication.
\begin{definition}\textbf{Cognitive semantic augmentation} is the enhancement of a cognitive agent's ability to interpret and process information by leveraging semantic knowledge. It enables the agent to prioritize relevant features, maintain semantic consistency, and improve the detection of important changes and patterns, thereby enhancing decision-making and understanding in complex environments.
\end{definition}
\subsubsection{End-to-end semantic-augmented communication}
As illustrated in Fig. \ref{2SatSEM}, the end-to-end communication between EO Sat 1, EO Sat 2, and the User Terminal (UT) involves several advanced steps to optimize the transmission and interpretation of EO data using the concept of Cognitive Semantic Augmentation (CSA). EO Sat 1 initiates the process by collecting signals from the Earth's surface, which could be in the form of imagery, radar, or other sensor data. Rather than transmitting raw data, Sat 1 applies DT-JSCC to compress the data into quantized semantic features, retaining only the most essential information needed for specific tasks, thus reducing communication overhead. These semantic features are transmitted over an inter-satellite link to EO Sat 2, which is enhanced with semantic processing capabilities. Upon receiving the data, EO Sat 2 conducts on-board semantic inference and augmentation processing, analyzing and further interpreting the features to extract meaningful insights, such as detecting objects, patterns, or changes over time. The processed semantic data is then transmitted to the User Terminal (UT), which allows the UT to access this refined information. Before EO Sat 2 transmits its downlink data, UT identically received the Sat1 EO data $U_{sem}(t_0)$ to perform SA technique \cite{pu2024fine} and optimize DT-JSCC decoder $\textit{l}_{UT}$. Therefore, the classification networks $\textit{f}_{s2}$ 
and $\textit{l}_{UT}$ can be updated individually as semantic-enhanced DT-JSCC to facilitate the CSA downlink. As illustrated in Fig.\ref{SEMtxrx}, we summarize the proposed end-to-end CSA satellite communication in Algorithm \ref{alg:CSA Satcom}.
\ref{alg:CSA Satcom}.
\begin{algorithm}[!t]
    \caption{\textbf{End-to-end CSA Satellite communication}} 
    \label{alg:CSA Satcom}
    \textbf{Input:} Sat1 and Sat2 EO image $U(t_i)$: Height $H$, Width $W$, and $D$ spectral bands; Semantic information: batch size $B$, number of classes $C$, and feature dimension $A$; \\
    \textbf{Output:} UT EO inference;
    \begin{algorithmic}[1]    
        \For{$i=0,1,\cdots,n-1$}
            \State Sat1 processing and transmit to Sat2 and UT: $U_{sem}(t_i)=\textit{f}_{s_1}(U(t_i))$.
            \State Sat2 and UT receive $U_{sem}(t_i)$ and processing $U_{sem}(t_{i+1})=\textit{f}_{s_2}(U(t_{i+1}))$. 
            \State Meta learning($\textit{g}_{s_2}$,$\textit{f}_{s_2}$,$\textit{l}_{s_2}$) \cite{pu2024fine} to process: 
            \State $\upsilon_{t_{i+1}}^{\textit{g}_{s_2}}\rightarrow$Minimise $\textit{L}_{SA}(\textit{f}_{s_2};\textit{g}_{s_2}|U_{sem}(t_i))$ and Optimizing $\textit{g}_{s_2}$ by backpropagating $\theta_{\textit{g}_{s_2}}$. \Comment{Covariance prediction network of Sat2}
            \State Meta-learning($\textit{g}_{UT}$,$\textit{f}_{UT}$,$\textit{l}_{UT}$) \cite{pu2024fine} to process:
            \State $\upsilon_{t_{i+1}}^{\textit{g}_{UT}}\rightarrow$Minimise $\textit{L}_{SA}(\textit{f}_{UT};\textit{g}_{UT}|U_{sem}(t_i))$ and Optimizing $\textit{g}_{UT}$ by backpropagating $\theta_{\textit{g}_{UT}}$. \Comment{Execute line 5 in parallel for covariance prediction network of UT}.
            \State Sat2 transmit $U_{sem}(t_{i+1})$ to UT $\textit{l}_{UT}$ for EO inference.
        \EndFor
	\end{algorithmic} 
\end{algorithm}
Notably, lines 4 and 5 are carrying out the CSA training of $\textit{f}_{s_2}$ on Sat2 side and $\textit{l}_{UT}$ on UT side simultaneously with lines 6 and 7. Following the completion of the CSA training, the data is subjected to semantic reconstruction, processing, and analysis to provide useful insights. These insights are then sent to user applications via Semantic Feature Division Multiple Access (SFDMA) \cite{ma2024semantic}. These user applications can use the semantically processed EO data for a variety of purposes, including inter-UE (User Equipment) interference cancellation, which improves communication efficiency and signal clarity. The SFDMA network reduces interference through approximate orthogonal transmission and introduces the Alpha-Beta-Gamma formula for semantic communications, with simulations validating its effectiveness. This overall process demonstrates how integrating semantic cognition into satellite systems can drastically improve the efficiency of EO data transmission, interpretation, and utilization, minimizing unnecessary data loads and focusing on task-relevant information.

\section{Numerical Results}
\subsection{Channel model}
Let  $X$  denote the transmitted signal sequence and $Y$  denote the received signal sequence. The general form of the channel equation is $Y = H \cdot X + N$, where \( H \) is the channel gain matrix and \( N \) is the noise term. Below are specific formulations for different types of channels.
\begin{itemize}
    \item \textbf{AWGN Channel}: No fading, so $H = 1$ which results in $Y = X + N_{\text{AWGN}}$, where $N_{\text{AWGN}} \sim \mathcal{N}(0, \sigma^2)$ is Gaussian noise with variance $\sigma^2$.
    \item \textbf{LEO Rician Channel}: Incorporates Rician fading, doppler shift, and satellite-specific effects: $Y = H_{\text{LEO-Rician}} \cdot X + N_{\text{AWGN}}$, where $H_{\text{LEO-Rician}}$ is a Rician-distributed fading gain, accounting for scintillation and atmospheric losses summarized in  Table~\ref{tab:sim_parameters}.
    \item \textbf{LEO Rayleigh Channel}: The fading gain follows a Rayleigh distribution with LEO-specific noise characteristics: $Y = H_{\text{LEO-Rayleigh}} \cdot X + N_{\text{AWGN}}$.
\end{itemize}
\begin{table}[ht]
\centering
\caption{Simulation Parameters}
\begin{tabular}{|l|c|}
\hline
\textbf{Parameter}                      & \textbf{Value} \\ \hline
Carrier frequency                        & 28 GHz        \\ \hline
Satellite antenna gain                   & 35 dBi        \\ \hline
User antenna gain                        & 37 dBi        \\ \hline
Scintillation loss                       & 0.5 dB        \\ \hline
Atmospheric loss                         & 0.3 dB        \\ \hline
Rician factor                            & 2.8           \\ \hline
\end{tabular}
\label{tab:sim_parameters}
\end{table}
To simulate propagation and attenuation, the channel model incorporates both small- and large-scale fading; the path loss components are the large-scale fading between the satellite and UT \cite{adam2024diffusion}:
$P_{L_{\text{tot}}} = P_{L_b} + P_{L_g} + P_{L_s}$, in which the attenuation resulting from atmospheric gases is denoted by $P_{L_g}$, the attenuation resulting from ionospheric or tropospheric scintillation by $P_{L_s}$, and the basic path loss is represented by $P_{L_b}$. Furthermore, the ISL of free space may be understood as $ P_{L_{\text{ISL}}} = P_{L_{ib}} $, in which $P_{L_{ib}}$ denotes Sat1 and Sat2's fundamental path loss.  These route loss components are all expressed in decibels (dB). More specifically, the free space propagation and shadow fading of the signal are taken into account using the fundamental path loss model $P_{L_b}$. For a distance $d_{k,i}$ also known as slant range, the free space path loss ($P_{L_{\text{FS}}}$) in dB range in meters and frequency $f_c$ in GHz. Hence, the slant distance $d_{k,i}$ can be expressed as:

\begin{equation}
    d = \sqrt{R_E^2 \sin^2 \theta + r_m^2 + 2R_Er_m - 2R_E r_m \sin \theta},
\end{equation}
where $R_E = 6378 \, \text{km}$ is the radius of the Earth, $r_m$ and $\theta$ are the satellite altitude and its elevation angle, respectively. 

Thereby, the free space path loss ($P_{L_{\text{FS}}}$) can be calculated as: $P_{L_{\text{FS}}}(d, f_c) = 32.45 + 20 \log_{10}(f_c) + 20 \log_{10}(d)$,
Shadow fading (SF) is modeled by a log-normal distribution as $\mathcal{N}(0, \sigma^2_{SF})$ with zero-mean and $\sigma_{SF}$ standard deviation. The values of $\sigma^2_{SF}$ can be extracted from the 3GPP Release-15. Then, the path loss with and without shadow fading in dB units is modeled as $P_{L_b} = P_{L_{\text{FS}}}(d, f_c) + SF$ and $P_{L_{ib}} = P_{L_{\text{FS}}}(d_{ib}, f_c) $, where $d_{ib}$ represent the distance of Sat1 and Sat2. 

We consider the Rician model to express the channel between LEO satellite \( s1 \) and the UT \( z \) and ISL between LEO satellite \( s1 \) and \( s2 \). The channel from the LEO satellite \( s1 \) to the UT \( z \)  and ISL are given as 

\begin{equation}
f_{s1,z} = \sqrt{\frac{R_z \zeta_{s1,z}}{R_z + 1}} \bar{f}_{s1,z} + \sqrt{\frac{\zeta_{s1,z}}{R_z + 1}} \tilde{f}_{s1,z},
\end{equation}
\begin{equation}
f_{s1,s2} = \sqrt{\frac{R_{s2} \zeta_{s1,s2}}{R_{s2} + 1}} \bar{f}_{s1,s2},
\end{equation}
where \( \bar{f}_{s1,z} \), \( \bar{f}_{s1,s2} \)  and \( \tilde{f}_{s1,z} \) are the line-of-sight (LoS) and non-line-of-sight (NLoS) paths components, respectively,  \( R_z \) is the Rician factor, and the ISL does not have NLoS component. After accounting for the LEO satellite antenna gain $G_T$, the large-scale fading effects of the LEO satellite $s1$ to the UT $z$ and ISL may be modeled as follows: $\zeta_{s1,z} (\text{dB}) = PL_{\text{tot}} - G_T$ and $\zeta_{s1,s2} (\text{dB}) = PL_{\text{ISL}} - G_T$. Furthermore, the complex Gaussian distribution with zero mean and unit variance is followed by the independent and identically distributed (i.i.d.) entries of \(\tilde{f}_{s1,z} \) data. 

The complex baseband downlink space domain channel at a time instance \( t \) and frequency \( f \) can be expressed as follows using a ray-tracing-based channel modeling method:
\begin{equation}
H_{s1,z}(t,f) =  f_{s1,z} \exp\left(2\pi \left[ t v_z - f \tau_z \right] \right),
\end{equation}
\begin{equation}
H_{s1,s2}(t,f) =  f_{s1,s2} \exp\left(2\pi \left[ t v_{s2} - f \tau_{s2} \right] \right),
\end{equation}
where  \( v_z \), \( v_{s2} \) are the Doppler shift, and \( \tau_z \), \( \tau_{s2} \) are the propagation delay. 
\subsection{Physical layer transmission}
\begin{figure}[!ht]
    \centering
    \includegraphics[scale=0.6]{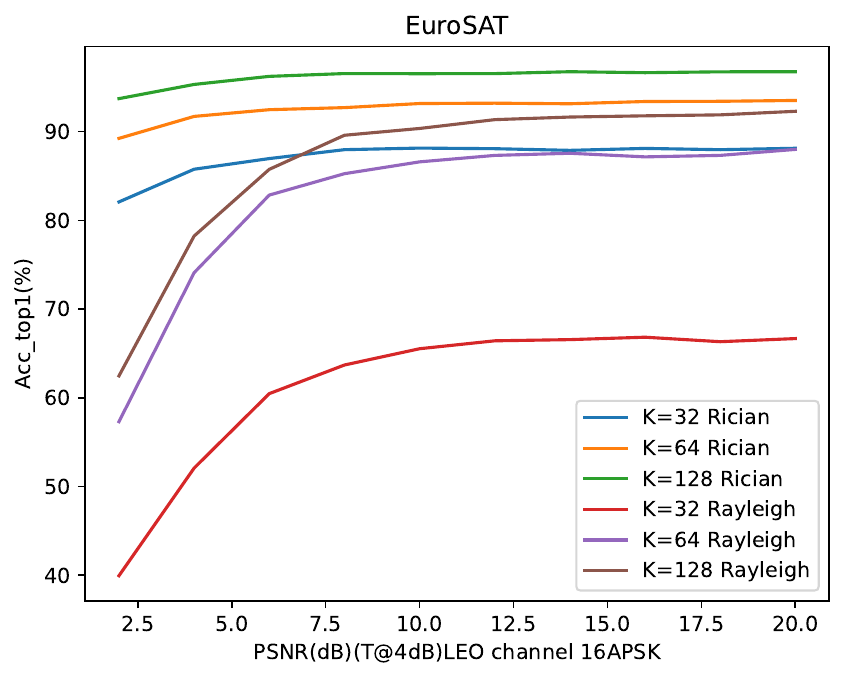}
    \caption{Top1 accuracy using DT-JSCC based on 16APSK LEO Rician and LEO Rayleigh channel.}
    \label{fig:top1_2}
\end{figure}

\begin{figure}[!ht]
    \centering
    \includegraphics[scale=0.38]{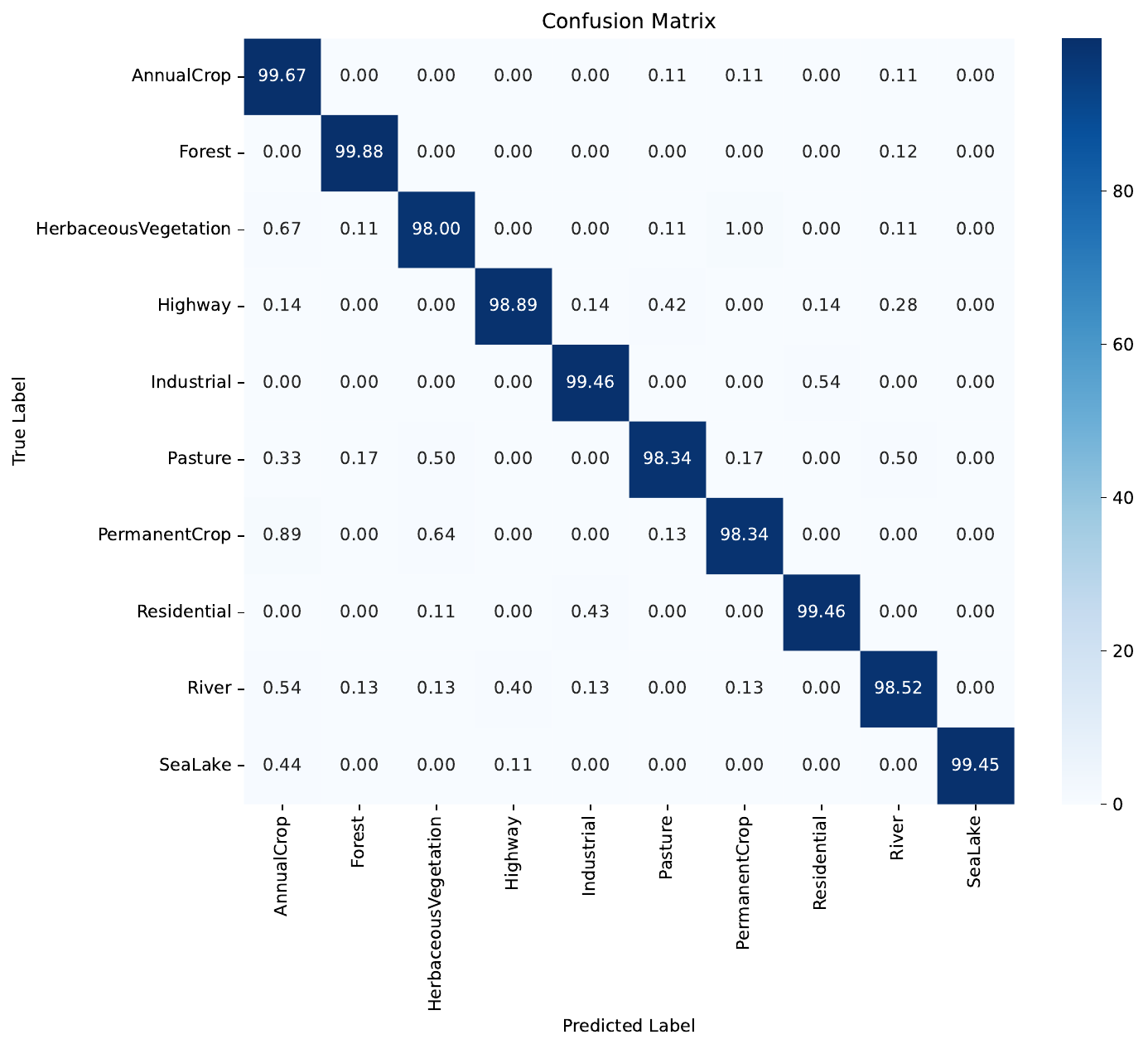}
    \caption{Top1 accuracy of confusion matrix using DTJSCC based on 16APSK Rician channel where PSNR=12dB and K=128.}
    \label{fig:cm_rician}
\end{figure}

Fig. \ref{fig:top1_2} presents Top-1 accuracy using DT-JSCC under different channel conditions for EuroSAT data. This graph depicts performance across different PSNR (Peak Signal-to-Noise Ratio) levels for the 16PSK modulation scheme. We show accuracy for the AWGN (Additive White Gaussian Noise) and Rician channels, with different K-factor values (32, 64, 128), where a higher K-factor generally corresponds to better channel quality. Furthermore, we compare accuracy across the LEO  Rician and LEO Rayleigh channels for certain K values, showing higher accuracy in the Rician channel.  Across all scenarios, the graphs indicate that increasing PSNR improves accuracy, with Rician channels generally performing better than Rayleigh, and higher K-factors leading to more stable performance, especially in LEO environments.
As illustrated in the confusion matrices of Fig.\ref{fig:cm_rician}, we can observe some key performance differences for DT-JSCC using K=128, the classification accuracy across most categories is notably high, with certain categories like "Forest" showing a strong prediction rate of 99.88\%, while there are minor misclassifications in categories like "PermanentCrop" and "Pasture" with accuracies of 98.34\%.
\subsection{CSA satellite network}
Table \ref{table:SA} shows CSA DT-JSCC system with high accuracy and evidently with most classes, such as Annual Crop (94.51\%), Herbaceous Vegetation (91.43\%), and Sea/Lake (98.85\%), showing high Top1 accuracy, indicating correct classifications. Some misclassifications can be improved, such as between Industrial and Residential areas. The non-CSA shows slightly reduced accuracy, with some increased misclassifications, particularly in Herbaceous Vegetation (89.64\%) and Sea/Lake (96.63\%). Overall, the CSA-enhanced system demonstrates 16\% accuracy improvement and fewer misclassifications than the non-CSA system.
\begin{table}[ht]
\centering
\caption{Top1 accuracy comparison of CSA and non-CSA  DT-JSCC K=32 systems while PSNR=12dB and 16APSK over LEO Rician channel.}
\begin{tabular}{|c|c|c|}
\hline
Class             & Accuracy (\%) CSA  & Accuracy (\%) non-CSA \\ \hline
AnnualCrop        & 94.50                        & 88.38                        \\ \hline
Forest            & 98.41                        & 98.84                        \\ \hline
HerbaceousVegetation & 91.63                     & 89.64                        \\ \hline
Highway           & 92.30                        & 77.27                        \\ \hline
Industrial        & 95.95                        & 90.95                        \\ \hline
Pasture           & 94.59                        & 90.53                        \\ \hline
PermanentCrop     & 90.03                        & 83.55                        \\ \hline
Residential       & 98.02                        & 95.95                        \\ \hline
River             & 92.33                        & 72.63                        \\ \hline
SeaLake           & 98.95                        & 98.33                        \\ \hline
\end{tabular}
\label{table:SA}
\end{table}

\begin{figure}[!ht]
    \centering
    \includegraphics[scale=0.6]{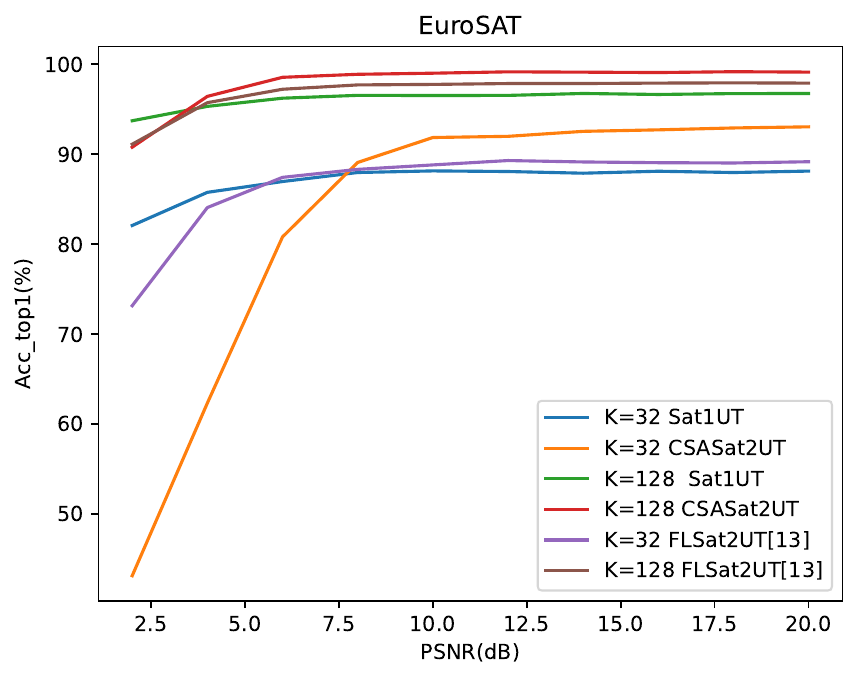}
    \caption{Top1 accuracy of CSA satellite networks using  DT-JSCC over 16APSK LEO Rician channel while DT-JSCC training at 4dB and CSA/federated learning training at 12dB.}
    \label{table:SA_Top1}
\end{figure}

As illustrated in Fig. \ref{table:SA_Top1}, the SA technique struggles significantly under the conditions of a low PSNR value due to the inherent noise and interference present in LEO Rician channels. This adverse environment leads to diminished signal quality, which adversely impacts the performance of the SA method. In contrast, a notable enhancement in Top1 accuracy is evident with both K=32 and K=128 DT-JSCC transmissions. This improvement can be attributed to the end-to-end CSA architecture, which leverages advanced techniques to optimize signal processing, resulting in a more robust transmission. By intelligently aggregating information, the system can better maintain signal integrity, thereby improving overall accuracy in challenging channel conditions. Given that federated learning aggregates updates from multiple clients on a central server and avoids transmitting raw data to protect privacy, our proposed end-to-end CSA scenario can analyze the semantic structure of source data from the nearby satellite and show its superiority over federated learning by leveraging 100-round communication between the UT server and client satellites, as presented in \cite{razmi2024board}. 
The results highlight the importance of employing sophisticated transmission strategies, particularly in environments characterized by high levels of noise and interference. The substantial gains in accuracy achieved with the DT-JSCC and CSA approaches underscore the efficacy of incorporating semantic-aware mechanisms in the design of communication systems, suggesting that further exploration and development in this area could yield even more significant benefits in future applications.

\section{Conclusion}
This work presents an advanced framework for semantic communication networks, with a focus on the development and optimization of cognitive semantic EO systems. The integration of semantic data augmentation, inter-satellite links, and cognitive processing capabilities ensures that only the most relevant, task-specific data is transmitted, significantly reducing communication overhead and improving system performance. Furthermore, by leveraging semantic inference and cognitive augmentation techniques, the system facilitates better image analysis, change detection, and decision-making in real-time applications. The proposed end-to-end model demonstrates how integrating semantic cognition into satellite systems can drastically improve data transmission, interpretation, and overall system efficiency, thereby addressing the demands of 6G networks and beyond. This research lays the groundwork for future advancements in semantic communication, with promising applications in a wide range of fields, including remote sensing, autonomous systems, and smart communication networks.

\section*{Acknowledgment}

This work was funded by the Luxembourg National Research Fund (FNR), with the granted SENTRY project corresponding to grant reference C23/IS/18073708/SENTRY.


\balance
\begin{IEEEbiographynophoto}{Hong-fu Chou} received the M.S. and Ph.D. degree in communication engineering from National Taiwan University, Taipei, Taiwan in 2006 and 2013, respectively. He is a Postdoctoral Fellow with the School of Computer Science, The University of Auckland from 2016 to 2020 and currently with Interdisciplinary Centre for Security, Reliability and Trust (SnT), University of Luxembourg. 
\end{IEEEbiographynophoto}
\begin{IEEEbiographynophoto}{Vu Nguyen Ha}  received the Ph.D. degree from the Institut National de la Recherche Scientifique-Énergie, Matériaux et Télécommunications (INRS-EMT), Université du Québec, Montréal, Québec, Canada, in 2017.  He is currently a Research Scientist of currently with Interdisciplinary Centre for Security, Reliability and Trust (SnT), University of Luxembourg. 
\end{IEEEbiographynophoto}
\begin{IEEEbiographynophoto}{Prabhu Kaliyammal Thiruvasagam} received his Ph.D degree in computer science and engineering from the Indian Institute of Technology Madras (IIT Madras). Currently, he is a Research Associate at SIGCOM group of SnT at the University of Luxembourg. His research interests include non-terrestrial networks, edge computing, joint sensing and communication, semantic communication, and sustainability aspects of future networks.
\end{IEEEbiographynophoto}
\begin{IEEEbiographynophoto}{Geoffrey Eappe} received the M.S. and Ph.D. degree in Vellore Institute of Technology University, Indore, India, in 2017 and 2021, respectively. He is a Postdoctoral Fellow with the Interdisciplinary Centre for Security, Reliability and Trust (SnT), University of Luxembourg. 
\end{IEEEbiographynophoto}
\begin{IEEEbiographynophoto}{Thanh-Dung Le} received a B.Eng. degree in mechatronics engineering from Can Tho University, Vietnam, an M.Eng. degree in electrical engineering from Jeju National University, S. Korea, and a Ph.D. in electrical engineering (Major in Applied Artificial Intelligence) from Ecole de Technologie Superieure (ETS), University of Quebec, Canada. From October 2023 to May 2024, he was a Postdoctoral Fellow with the Biomedical Information Processing Laboratory, ETS. Before that, he joined the Institut National de la Recherche Scientifique, Canada, where he researched classification theory and machine learning. He is currently a Research Associate at the Interdisciplinary Center for Security, Reliability, and Trust at the University of Luxembourg. 
\end{IEEEbiographynophoto}
\begin{IEEEbiographynophoto}{Nguyen Ti Ti} received the B.Eng. degree in electrical engineering from Ho Chi Minh City University of Technology, Vietnam, in 2013, the M.Eng. degree in embedded system from University of Rennes 1, France, in 2015, and the Ph.D. degree in telecommunications at the Institut National de la Recherche Scientifique (INRS), Universit\'{e} du Qu\'{e}bec, Canada, 2020. From 2020-2024, he was a postdoc fellow in Synchromedia - Ecole de Technologie Sup\'{e}rieure, Universit\'{e} du Qu\'{e}bec.  He is currently a Research Associate at the Interdisciplinary Centre for Security, Reliability, and Trust, University of Luxembourg. 
\end{IEEEbiographynophoto}
\begin{IEEEbiographynophoto}{Duc-Dung Tran} received the B.Eng. degree in electronics and telecommunications from Hue University of Sciences, Vietnam, in 2013, the M.Sc. degree in computer sciences from Duy Tan University, Vietnam, in 2016, and the Ph.D. degree in computer sciences from University of Luxembourg in 2024. From 2014 to 2019, he was with the Faculty of Electrical and Electronics Engineering, at Duy Tan University. He is currently a Research Associate at the Interdisciplinary Center for Security, Reliability and Trust (SnT), University of Luxembourg. His current research interests include 5G and beyond wireless networks, URLLC, multiple access techniques, and machine learning for terrestrial and satellite communications. 
\end{IEEEbiographynophoto}
\begin{IEEEbiographynophoto}{Luis Manuel GARCES-SOCARRAS} received his Ph.D degree in electrical and electronic engineering from the Technological University of Havana José Antonio Echeverría / Universidad Tecnológica de La Habana José Antonio Echeverría (CUJAE). Currently, he is a Research Associate at SIGCOM group of SnT at the University of Luxembourg. 
\end{IEEEbiographynophoto}
\begin{IEEEbiographynophoto}{Juan Carlos MERLANO-DUNCAN} (Senior
Member, IEEE) received the Diploma degree in
electrical engineering from the Universidad del
Norte, Barranquilla, Colombia, in 2004, and the
M.Sc. and Ph.D. Diploma degrees (cum laude)
from the Universitat Politècnica de Catalunya
(UPC), Barcelona, Spain, in 2009 and 2012,
respectively.
At UPC, he was responsible for the design
and implementation of a radar system known as
SABRINA, which was the first ground-based bistatic radar receiver using
space-borne platforms, such as ERS-2, ENVISAT, and TerraSAR-X as
opportunity transmitters (C and X bands). He was also in charge of the implementation of a ground-based array of transmitters, which was able to monitor
land subsidence with subwavelength precision. These two implementations
involved FPGA design, embedded programming, and analog RF/Microwave
design. In 2013, he joined the Institute National de la Recherche Scientifique,
Montreal, QC, Canada, as a Research Assistant in the design and implementation of cognitive radio networks employing software development and
FPGA programming. He has been with the University of Luxembourg, since
2016, where he currently works as a Research Scientist with the COMMLAB
Laboratory working on SDR implementation of satellite and terrestrial communication systems and passive remote sensing systems. His research interests include wireless communications, remote sensing, distributed systems,
frequency distribution and carrier synchronization systems, software-defined
radios, and embedded systems.
\end{IEEEbiographynophoto}
\begin{IEEEbiographynophoto}{Symeon Chatzinotas} (Fellow, IEEE) is currently Full Professor / Chief Scientist I and Head of the SIGCOM Research Group at SnT, University of Luxembourg. He received the M.Eng. degree in telecommunications from the Aristotle University of Thessaloniki, Thessaloniki, Greece, in 2003, and the M.Sc. and Ph.D. degrees in electronic engineering from the University of Surrey, Surrey, U.K., in 2006 and 2009, respectively. 
He has (co-)authored more than 400 technical papers in refereed international journals, conferences and scientific books. He is currently in the editorial board of the IEEE Open Journal of Vehicular Technology and the International Journal of Satellite Communications and Networking.
\end{IEEEbiographynophoto}

\end{document}